\documentclass[%
 reprint,
superscriptaddress,
showkeys,
 amsmath,amssymb,
 aps,
]{revtex4-1}

\usepackage{graphicx}
\usepackage{dcolumn}
\usepackage{bm}

\usepackage{ulem}

\usepackage{float}

\newcommand{\note}[1]{\marginpar{\tiny {#1}}}   
\newcommand{\SKIP}[1]{\note{SKIP'D}}
\newcommand{\TKE}{{\rm TKE}}		
\newcommand{\TXE}{{\rm TXE}}		
\newcommand{\cchi}{\mbox{\boldmath $\chi$}}
\newcommand{\nubar}{{\bar{\nu}}}	
\newcommand{\nuL}{{\bar\nu}_{\rm L}}	
\newcommand{\nuH}{{\bar\nu}_{\rm H}}	

\begin{document}

\preprint{APS/123-QED}

\title{Correlation studies of fission 
fragment neutron multiplicities}

\author{M. Albertsson}
\affiliation{Mathematical Physics, Lund University, S-221 00 Lund, Sweden}
\author{B.G. Carlsson}
\affiliation{Mathematical Physics, Lund University, S-221 00 Lund, Sweden}
\author{T. D{\o}ssing}
\affiliation{Niels Bohr Institute, University of Copenhagen, 2100 Copenhagen {\O}, Denmark}
\author{P. M{\"o}ller} 
\affiliation{Mathematical Physics, Lund University, S-221 00 Lund, Sweden}
\affiliation{P.\ Moller Scientific Computing and Graphics, Inc.,
		 P.O.\ Box 75009, Honolulu, HI 96836 USA}
\author{J. Randrup}
\affiliation{Nuclear Science Division, Lawrence Berkeley National Laboratory,
Berkeley, California 94720, USA}
\author{S. {\AA}berg}
\affiliation{Mathematical Physics, Lund University, S-221 00 Lund, Sweden}

\date{\today}

\begin{abstract}
We calculate neutron multiplicities from fission fragments with specified mass
numbers for events having a specified total fragment kinetic energy. 
The shape evolution from the initial compound nucleus to the scission 
configurations is obtained with the Metropolis walk method
on the five-dimensional potential-energy landscape,
calculated with the macroscopic-microscopic method
for the three-quadratic-surface shape family.
Shape-dependent microscopic level densities are used to guide the random walk,
to partition the intrinsic excitation energy between the two proto-fragments
at scission, and to determine the spectrum of the neutrons evaporated
from the fragments.
The contributions to the total excitation energy of the resulting fragments
from statistical excitation and shape distortion at scission is studied.
Good agreement is obtained with available experimental data on neutron 
multiplicities in correlation with fission fragments from $^{235}$U(n$_{\rm th}$,f).
At higher neutron energies a superlong fission mode 
appears which affects the dependence of the observables
on the total fragment kinetic energy.
\end{abstract}

\maketitle

\raggedbottom

\section{Introduction}
\label{intro}
A long-standing challenge in nuclear fission is the dependence 
of the average neutron multiplicity $\nubar$ on the fragment mass number $A$.
The characteristic sawtooth behavior of $\nubar(A)$ is well illustrated for 
$^{235}$U(n$_{\text{th}}$,f), see e.g.\ Ref.\ \cite{nishio236u}.
Because the number of neutrons evaporated is indicative of the excitation 
energy in the emitting fragment,
it is of key importance to understand the degree of excitation
of the fission fragments, as a function of $A$.

To a good approximation,
the total excitation energy of a given fission fragment is the sum of two 
distinct contributions.
One is the share of total statistical excitation 
received by the distorted proto-fragment at the time of scission.
The other contribution results from the relaxation of the fragment shape 
from its distorted form at scission to its equilibrium shape
which converts the change in potential energy into additional fragment heat.
Different theoretical descriptions of the fission process
\cite{Wilkins1976,Dubray2008,ScampsNature2018,
BulgacPRL116,Bulgac2018,Albertsson2020} 
have yielded different results for the amount of distortion energy.
Furthermore, because it is the sum of the two contributions
that determines the energy available for neutron evaporation, 
it is difficult to determine the individual contributions
from the measured $\nubar(A)$ alone.

However, by studying the dependence of $\nubar(A)$ on
the total fragment kinetic energy (\TKE) it may be possible to gain important 
insight into how the fission fragment excitation energy is composed.
Such correlation measurements were performed recently,
yielding $\nubar(A;\TKE)$ for $^{235}$U(n$_{\rm th}$,f) \cite{Gook2018}.
In this paper,
we discuss how variations of \TKE\ are associated with structure-dependent 
variations in the contributions to the fission fragment excitations,
leading in turn to observable variations of the 
\TKE-constrained neutron multiplicity, $\nubar(A;\TKE)$.

Due to the highly dissipative character of collective nuclear dynamics,
it has proven possible to model the shape evolution of a fissioning nucleus
as a Metropolis walk on the multi-dimensional potential-energy
surface~\cite{RandrupPRL106,RandrupPRC84,RandrupPRC88}.
By using shape-dependent microscopic level densities~\cite{UhrenholtNPA913} 
for guiding the shape evolution, 
a consistent framework was obtained for calculating the energy-dependent 
fission-fragment mass distribution \cite{Ward2017}.

However, in the region of symmetric fission
the scission configurations lead to too little statistical excitation \cite{Albertsson2020}
and to too high \TKE.
While this problem is of little importance as long as the focus is on the
fragment-mass yields (which tend to be very small in the symmetric region),
it is relevant for the calculation of $\nubar(A)$.
Therefore, in the present study of correlated neutron multiplicities,
we consider only fission events in the asymmetric region,
namely $A_{\rm L} \leq 104$ (and correspondingly $A_{\rm H} \geq 132$).

Recently, shape-dependent microscopic level densities
were employed also for the calculation of the excitation energy partition 
between the fission fragments \cite{Albertsson2020}.
In that treatment, it was assumed that the statistical excitation energy
available at scission is divided microcanonically between the two
proto-fragments whose distorted shapes 
later on relax to their ground-state forms.
It was found  \cite{Albertsson2020} that this treatment
leads to a reasonably good reproduction of $\nubar(A)$ 
measured for $^{235}$U(n,f) at both thermal energies
and $E_{\rm n}=5.55$\,MeV \cite{MullerPRC29}, 
in particular for asymmetric fission events.
The study brought out the important influence of the specific structure
of the various proto-fragments whose level densities affect the energy
partitioning significantly.

We now go further and study the energy dependence of the structure effects 
by gating on specific values of \TKE.
In such a more detailed study,
the specification of a particular \TKE\ value
selects the total excitation energy \TXE. The resulting
fragment excitation energies can then be calculated and the associated 
mean neutron multiplicities, $\nubar(A;\TKE)$, can be obtained.

Thus, for the first time, a fission model based on microscopic level densities, 
combined with a 5D potential-energy surface
 obtained with the macroscopic-microscopic method, 
is applied to calculate more complex correlation observables, 
namely the average neutron multiplicity from fission fragments of given $A$
for events with a particular \TKE.
A phenomenological deterministic model of prompt neutron emission was
recently applied to the same problem yielding very good agreement with data
\cite{Tudora2019}.

The method of the calculation is briefly presented in Sect.\ \ref{sec:method}. 
In Sect.\ \ref{sec:energies} we discuss contributions from intrinsic and
distortion energy at scission to the excitation energies of the 
primary fission fragments,
and in Sect.\ \ref{sec:correlation} the results 
for the neutron multiplicities are presented.
Finally, Sect.\ \ref{sec:summary} presents a summary and a discussion.

\section{Method of calculation}
\label{sec:method}
The calculations  closely follow Ref.\ \cite{Albertsson2020}.
The evolution of the nuclear shape from the ground-state shape
to scisson is treated as a Metropolis random walk 
on the potential-energy landscape $U(\cchi)$ \cite{RandrupPRL106}. 
The shape \cchi\ is described by the 3QS parametrization 
\cite{Nix1969,Moller2009} which has five parameters: 
the overall elongation given by the quadrupole moment $q_2$, 
the neck radius $c$, the spheroidal deformations 
$\varepsilon_{\mathrm{f1}}$ and $\varepsilon_{\mathrm{f2}}$ 
of the endcaps of the two nascent fragments, 
and the mass asymmetry $\alpha$.
For each of the more than 6 million nuclear shapes considered, 
the microscopic level density is calculated by the combinatorial method
\cite{UhrenholtNPA913} up to about 6 MeV of excitation energy
and is extrapolated to higher energies 
using the calculated shell and pairing energies \cite{Ward2017}.

The initial configuration is assumed to be a compound nucleus
having the excitation energy $E_0^*=E_{\rm n}+S_{\rm n}$,
where $E_{\rm n}$ is the kinetic energy of the incoming neutron being absorbed
and $S_{\rm n}$ is the corresponding neutron separation energy.
With $M_0$ as the mass of the compound nucleus,
the total energy is given by $E_{\rm tot}=M_0+E_0^*$ which is conserved
during the subsequent evolution. 
Consequently, at a given shape \cchi\, the local intrinsic energy
is given by $E^*(\cchi)=E_{\rm tot}-U(\cchi)$ because the shape motion is assumed to be 
so strongly damped that the local collective kinetic energy is negligible.

In the considered fission reaction $^{235}$U(n,f) the compound nucleus
$^{236}$U can have either angular momentum $I$ = 3 or 4. The angular momentum
is conserved in the Metropolis random walk by considering  level densities with
a fixed angular momentum for each shape. Since $I$ = 3 and 4 give very similar
result \cite{Ward2017} the presented calculations are performed at $I$ = 4.

As in our earlier work \cite{Ward2017},
the shape changes are selected by the Metropolis method
using the associated shape-dependent microscopic level densities 
$\rho(\cchi)$, ensuring detailed balance,
$P(\cchi\to\cchi')/P(\cchi'\to\cchi)=\rho(\cchi')/\rho(\cchi)$.

The asymmetry $\alpha$ is asssumed to be frozen in
when the neck radius has shrunk to $c=c_0=2.5$~fm \cite{Ward2017,Albertsson2020}.
Subsequently, the system reaches a scission configuration at $c=c_{\rm sc}=1.5$~fm
\cite{Albertsson2020}
where the shapes of the proto-fragments are determined 
(see Sect.\ \ref{sec:energies_2}),
and the available intrinsic energy is partitioned between them
(see Sect.\ \ref{sec:energies_1}).
The initially distorted proto-fragments are being accelerated 
by their mutual Coulomb repulsion
and the shapes eventually revert to equilibrium forms.
Their original distortion energies are thereby 
converted to additional intrinsic excitation.
Subsequently, after full acceleration has been achieved,
each excited primary fragment evaporates neutrons
as long as it is energetically possible (see Sect.\ \ref{sec:correlation}).

For each reaction case considered,
a total of $10^6$ fission events are generated and
for each one we record the mass numbers of the two primary fission fragments,
$A_{\rm L}$ and $A_{\rm H}$, their total kinetic energy \TKE,
as well as the number of neutrons evaporated from each one,
$\nu_{\rm L}$ and $\nu_{\rm H}$,
as would be done in an ideal experiment.

\section{Energies in the fission process}
\label{sec:energies}
In Sect.\ \ref{sec:energies_1} we discuss the various key energies,
namely the total fragment excitation energy (\TXE)
and the corresponding total fragment kinetic energy (\TKE),
as well as the decomposition of the individual proto-fragment excitations
into intrinsic and distortion energies.
Then, in Sect.\ \ref{sec:energies_2} we describe
how the intrinsic energy available at scission
is partitioned between the two proto-fragments and how this depends on \TKE. 
Finally, in Sect.\ \ref{sec:energies_3} we analyze
the variation of the distortion energy with \TKE\ and fragment mass.

\subsection{Key energy quantities}
\label{sec:energies_1}

Once the initial compound nucleus has been prepared,
we follow an ensemble of shape evolutions,
as described in Sect.\ \ref{sec:method}.
These represent possible evolutions of the fissioning system
subject to the conservation of the total energy $E_{\rm tot}$.
When the evolving system has attained its scission shape, $\cchi_{\rm sc}$,
it is assumed to divide into two distorted and excited proto-fragments
which subsequently recede and accelerate 
while their shapes gradually relax to their equilibrium forms.

We assume that the strongly damped description of the shape evolution
remains valid until scission, 
so the proto-fragments are formed with vanishing kinetic energy.
Furthermore, the fragments typically carry several ($\approx$4) units 
of angular momentum,
but we ignore the associated rotational energy which is relatively small.
The combined intrinsic excitation energy of the two proto-fragments
at scission is then given by the corresponding local intrinsic energy,
\begin{equation}
E_{\rm L}^{\rm intr} +E_{\rm H}^{\rm intr} 
=E^*(\cchi_{\rm sc}) =E_{\rm tot}-U(\cchi_{\rm sc})\ .
\end{equation}
The partitioning of the intrinsic energy among the two fragments
is assumed to be statistical (see Sect.\ \ref{sec:energies_2}).

For a given mass partition, $A_0 \to A_{\rm L}+A_{\rm H}$,
the $Q$ value is $Q_{\rm LH}^* =  M_0 + E_0^* - M_{\rm L} - M_{\rm H}$.
This quantity represents the energy available
for the total fragment kinetic energy, \TKE,
and the combined excitation energies of the two primary fragments, \TXE,
after the acceleration and the shape relaxation have been completed,
\begin{equation}\label{TT}
Q_{\rm LH}^* = \TKE\ + E_{\rm L}^* + E_{\rm H}^* = \TKE + \TXE\ .
\end{equation}

In the present study, we assume that the proto-fragments
have spheroidal shapes with eccentricities equal to those of the
corresponding endcaps of the 3QS scission shape,
$\varepsilon_{\rm L,sc}=\varepsilon_{\rm f1}(\cchi_{\rm sc})$ and
$\varepsilon_{\rm H,sc}=\varepsilon_{\rm f2}(\cchi_{\rm sc})$.
The distortion energy of proto-fragment $i={\rm L,H}$ can then be expressed as
$E^{\rm dist}_i=U_i(\varepsilon_{i,\rm sc})-U_i(\varepsilon_{i,\rm gs})$,
where $U_i(\varepsilon)$ denotes the potential energy of deformation of the fragment
(see Sect.\ \ref{sec:energies_3}).
As the proto-fragments gradually attain their ground-state shapes,
their distortion energies are being converted into 
additional intrinsic excitation energy,
so the final excitation energy is the sum of 
the original intrinsic energy at scission and the distortion energy,
\begin{equation}
E^*_{\rm L} = E^{\rm intr}_{\rm L}+E^{\rm dist}_{\rm L}\ ,\
E^*_{\rm H} = E^{\rm intr}_{\rm H}+E^{\rm dist}_{\rm H}\ .
\end{equation}
Thus, with $\TXE=E^*_{\rm L}+E^*_{\rm H}$,
the total kinetic energy is determined from Eq.\ (\ref{TT}),
$\TKE=Q^*_{\rm LH}-\TXE$.

\begin{figure}[t]
\centering
\includegraphics[width=1.0\linewidth]{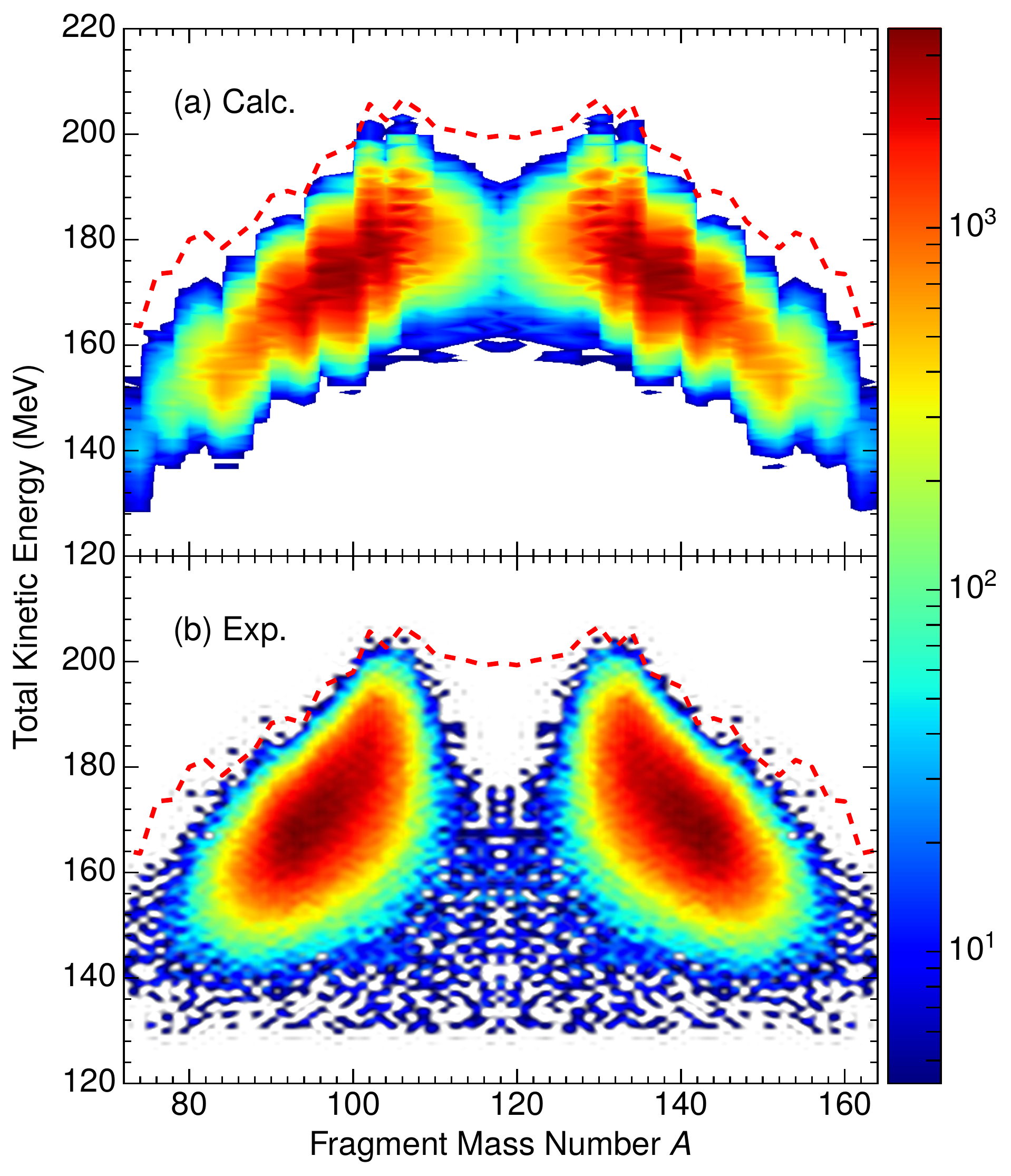}
\caption{
Number of scission events in log-scale versus 
fission fragment mass number and total kinetic energy
for $^{235}$U($n_{\text{th}},f)$. In (a) calculated results
with $c_0$=1.5 fm, and in (b) experimental data from Ref.\ \cite{Al2013}.
The dashed curve shows $Q$ values for different fragment masses. 
Experimental number of events are scaled to the same number of
events as calculated, namely $10^6$.
}
\label{fig:TKEvsA} 
\end{figure}

Figure \ref{fig:TKEvsA} shows contour plots of the calculated (a) 
and measured (b) number of scission events with respect to fragment mass number $A$
and total fragment kinetic energy \TKE\ for $^{235}$U(n$_{\text{th}},f)$.

In the region of symmetric fission,
the measured \TKE\ values are considerably lower than the calculated values.
This is probably caused by too small elongations of the calculated fission 
configurations at scission that leads to too small statistical excitation 
and an overestimation of the \TKE.
Due to this problem, we restrict the present study to asymmetric fission.

In the asymmetric region ($A_{\rm L}\leq104$ and $A_{\rm H}\geq132$),
the measured most probable \TKE\ shows a gradual decrease 
with increasing asymmetry. This feature is well reproduced by the calculations.

However, the width of the \TKE\ distribution for a given $A$
is underestimated in the calculation by typically 20\%.
This may (at least in part) be due to the fact that the calculations 
include only even-even fragment pairs having (approximately) equal $N/Z$ ratios,
namely that of $^{236}$U.
As a consequence of this restriction,
there is only one $(N,Z)$ combination for a given $A$,
whereas the actual fission process populates several combinations
and thus leads to a broader \TKE\ distribution.

The fixed $(N,Z)$ combination for given fragment mass number $A$, assumed in the calculations,
implies a well defined $Q$ value, shown by a dashed curve in Fig.\ \ref{fig:TKEvsA}a. This constitutes 
the maximum possible value of \TKE\, $\TKE_{\rm max}(A)=Q^*(A)$,
and the bulk of the events lie well below this boundary.
In experiments different $(N,Z)$ combinations are possible for each $A$,
leading to a set of $Q$ values. However, in Fig.\ \ref{fig:TKEvsA}b we show the $Q$ values 
from Fig.\ \ref{fig:TKEvsA}a.

For a given mass asymmetry (and thus a given $Q$ value),
the variation of \TKE\ is counterbalanced by the variation of \TXE,
the combined excitation of the primary fragments.
Consequently, by gating on \TKE\
it is possible to investigate a \TXE\ range of about 10-40 MeV,
for a given asymmetry.
In particular, if, for a given mass partition,
the specified \TKE\ value is gradually decreased, 
starting from the maximally realized value, 
the available \TXE\ will exhibit a corresponding gradual increase.
This, in turn, will be reflected in the mean number of neutrons 
evaporated from each of the two fragments, 
$\nubar_{\rm L}$ and $\nubar_{\rm H}$,
which will provide more detailed experimental information on 
the origin of $E^*_{\rm L}$ and $E^*_{\rm H}$.

\begin{figure}[t]
\centering
\includegraphics[width=1.0\linewidth]{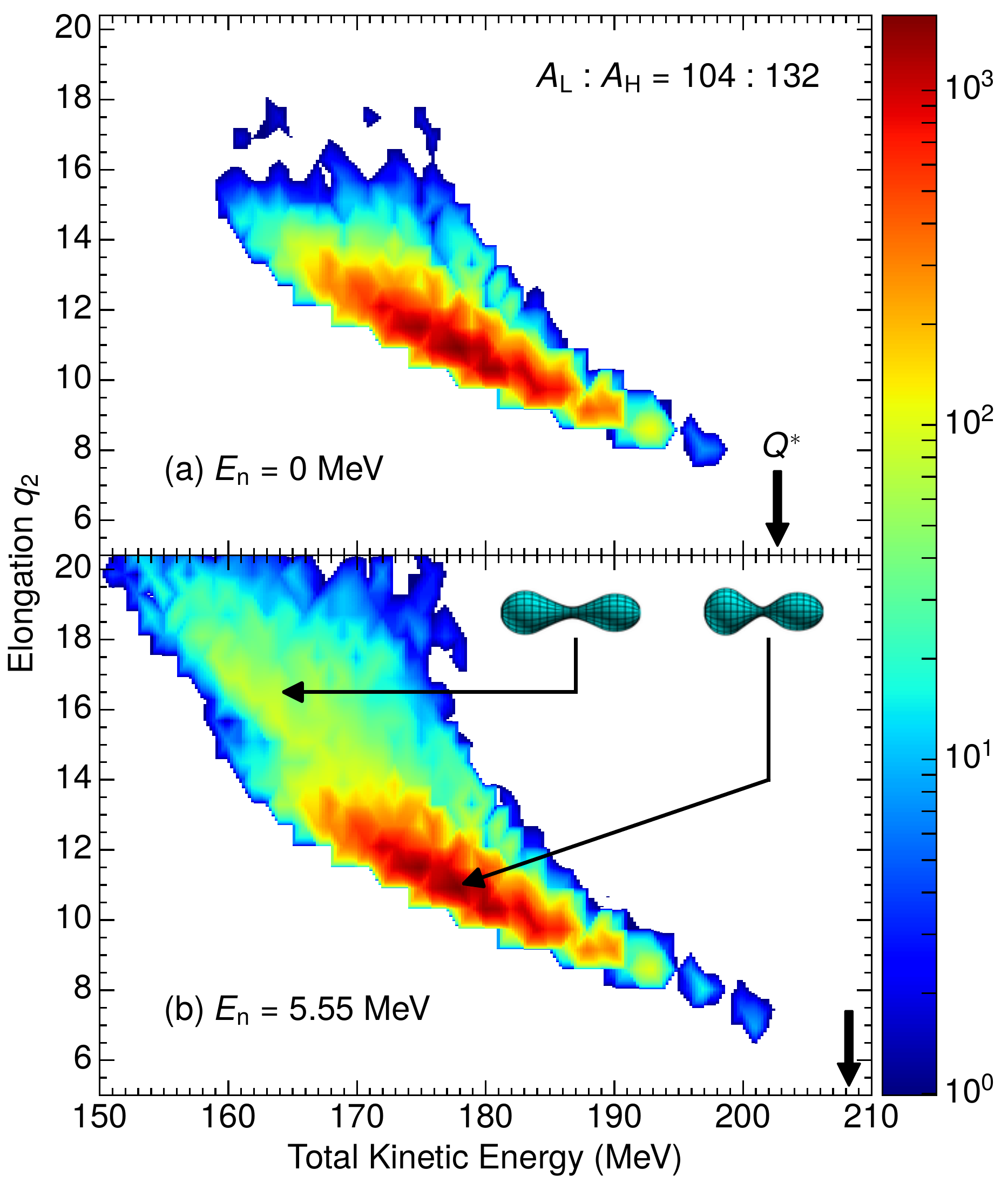}
\caption{Number of scission events in log-scale for fragment mass
division $A_{\text{L}}$:$A_{\text{H}}$=104:132 versus total kinetic energy \TKE\
and elongation $q_2$. In (a) $E_{\text{n}}$=0 and in (b) $E_{\text{n}}$=5.55 MeV.
In (b) typical shapes are shown for the superlong ($q_2\approx 16.5$)
and standard ($q_2\approx11$) modes. Arrows indicate 
the $Q^*$-values.
}
\label{fig:q2vsTKE} 
\end{figure}

In general, a fission event having a small \TKE\ value
(i.e.\ a large \TXE\ value)
is associated with a rather elongated scission configuration,
as measured for example by the quadrupole moment 
of the density distribution, $q_2$.
This feature is illustrated quantitatively in Fig.\ \ref{fig:q2vsTKE}
showing contour plots of the number of scission events versus
\TKE\ and $q_2$ for the mass split,
$A_{\rm L}:A_{\rm H}=104:132$ for $E_{\text{n}}$=0 (a)
and $E_{\text{n}}$=5.55 MeV (b).
At $E_{\text{n}}$=0 a wide range of quadrupole moments occur,
$8 < q_2 < 18 $, for the considered mass partition.
The most compact scission shapes (having the smallest $q_2$ values)
are associated with large \TKE\ values close to the $Q$ value,
while the most elongated scission shapes
(having the largest $q_2$ values)
have \TKE\ values that are about 40 MeV smaller.
The scission shapes are thus strongly dependent
on the considered \TKE\ value. 

In Fig.\ \ref{fig:q2vsTKE}b we show that even more elongated
scission shapes can occur when more energy is made available
by increasing the kinetic energy of the incident neutron. This is 
related to the appearance of a superlong fission mode.
In 4-D Langevin calculations \cite{Usang2019} the symmetric fission
events at low \TKE\ (140-170 MeV) seen in Fig. \ref{fig:TKEvsA}b, 
valid for thermal neutrons,
could be related to the superlong mode. With increasing excitation
energy this fission mode becomes important also for asymmetric
fission, that is discussed in Sect.\ \ref{sec:Higher}.

\subsection{Intrinsic excitation energies}
\label{sec:energies_2}

\begin{figure}[t]
\centering
\includegraphics[width=1.0\linewidth]{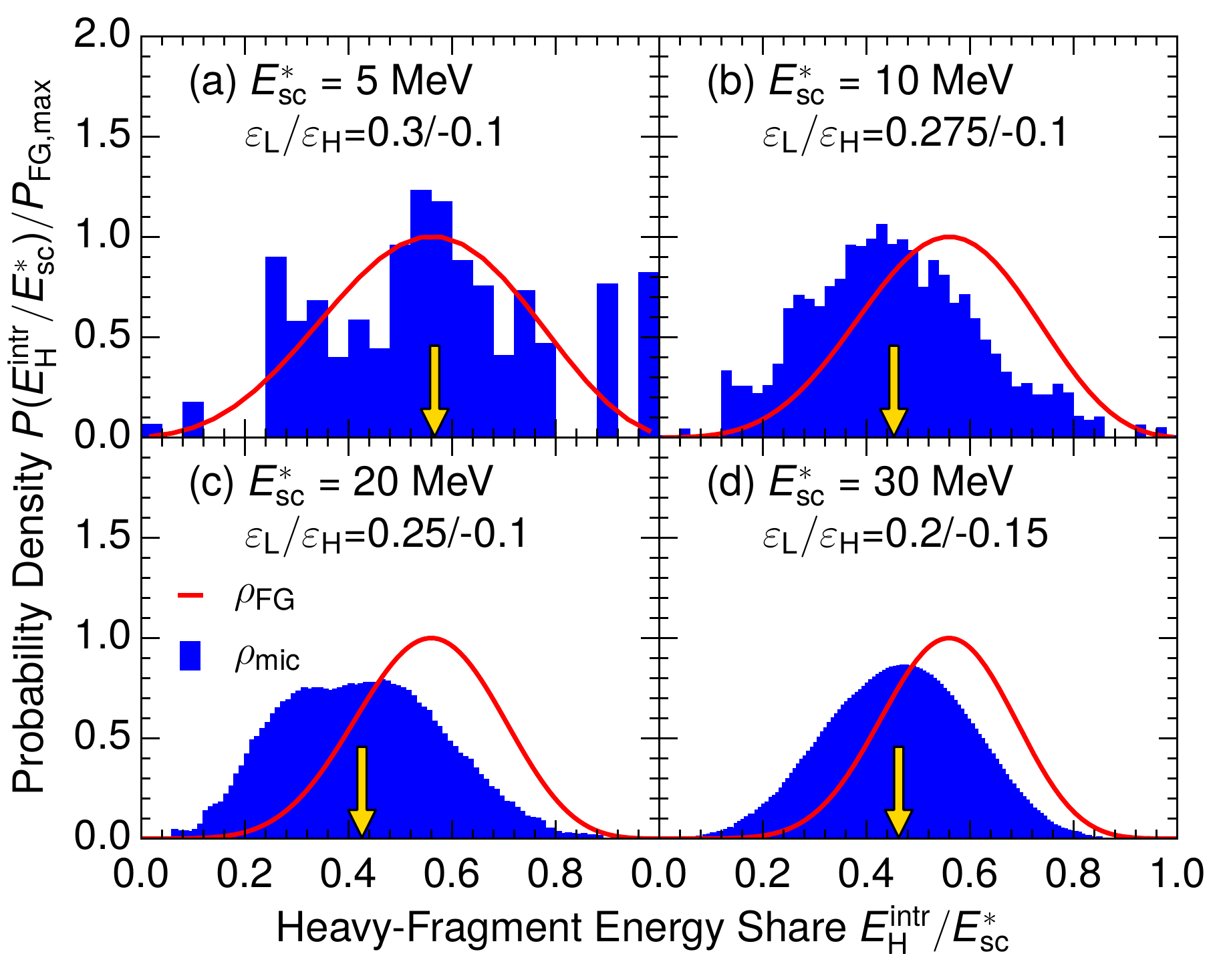}
\caption{Probability density for the heavy-fragment share $E^{\rm intr}_{\text{H}}/E^*_{\text sc}$ for different
intrinsic energies $E^*_{\text{sc}}=E^{\rm intr}_{\text{L}}+E^{\rm intr}_{\text{H}}$ for the fragment
mass division $A_{\text{L}}$:$A_{\text{H}}$=104:132.
The blue histogram
is obtained utilizing microscopic level densities and the red curve utilizing Fermi-gas level densities.
In (a) $E_{\text{sc}}^*=$5 MeV, (b) $E_{{\text{sc}}}^*=$10 MeV, (c) $E_{\text{sc}}^*=$20 MeV, 
and (d) $E_{\text{sc}}^*=$30 MeV. 
Deformations of heavy and light fragments, $\varepsilon_{\rm L}/\varepsilon_{\rm H}$, are average values at given
total excitation energy.
Arrows mark the average energy from microscopic level densities.
}
\label{fig:ProbDens3} 
\end{figure}

\begin{figure}[t]
\centering
\includegraphics[width=1.0\linewidth]{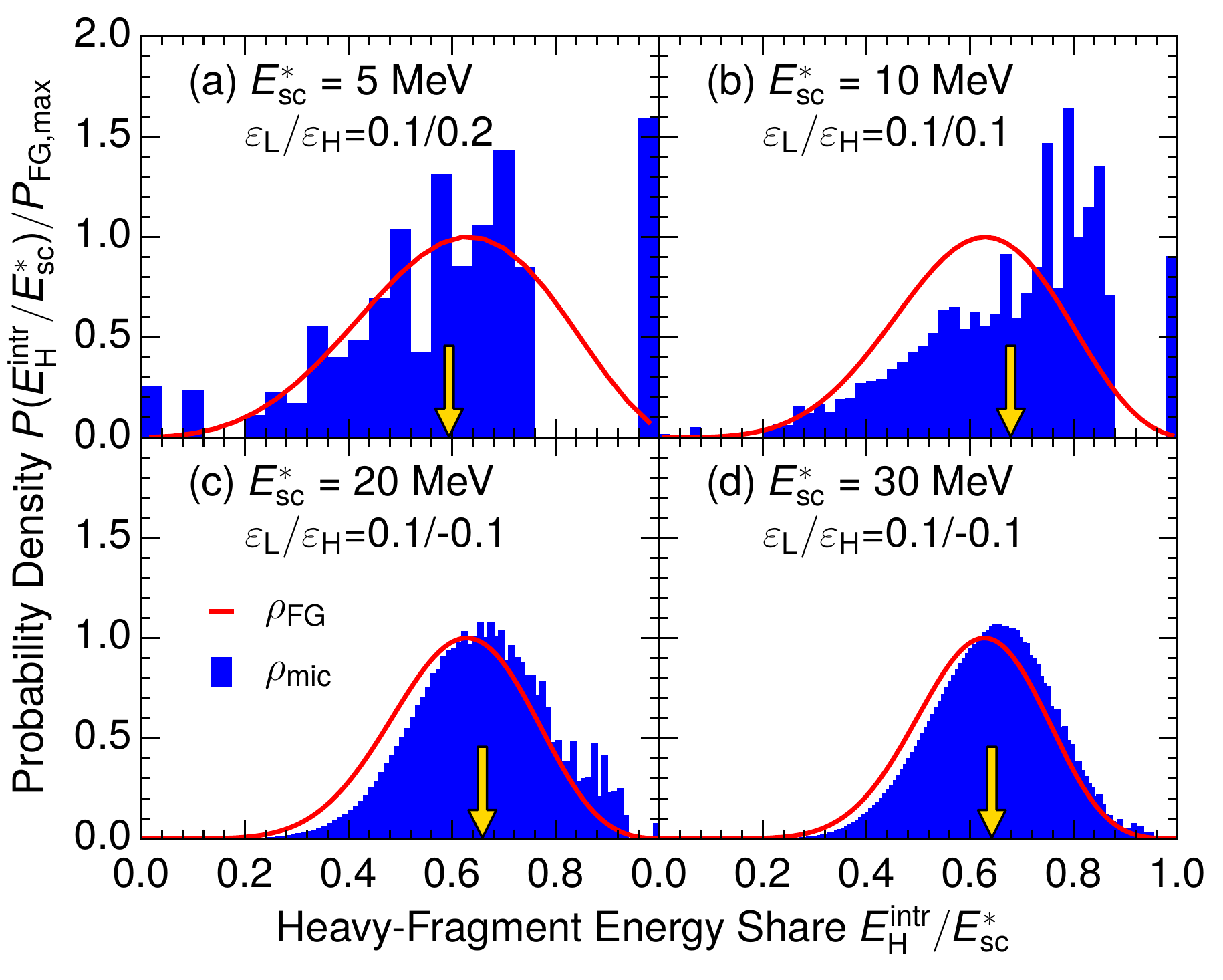}
\caption{
Similar to Fig.\ \ref{fig:ProbDens3} but for $A_{\text{L}}$:$A_{\text{H}}$=88:148.
}
\label{fig:ProbDens4} 
\end{figure}

We assume that the intrinsic excitation energy available at scission,
$E^*_{\rm sc}$, is divided statistically between the two proto-fragments,
as proposed in Ref.\ \cite{Albertsson2020}.
Thus the probability distribution for the intrinsic excitation
of the heavy fragment, $E^{\rm intr}_{\rm H}$, has a microcanonical form,
\begin{equation}
P(E^{\rm intr}_{\rm H};E^*_{\rm sc}) \sim 
\rho_H(E^{\rm intr}_{\rm H};\varepsilon_{\rm H,sc})\,
\rho_L(E^*_{\rm sc}-E^{\rm intr}_{\rm H};\varepsilon_{\rm L,sc}),
\label{eq:partition}
\end{equation}
where $\rho_{\rm H}(E^*)$ and $\rho_{\rm L}(E^*)$ 
are the effective level densities of the heavy and light proto-fragments 
evaluated at their respective shapes at scission.
Figures \ref{fig:ProbDens3} and \ref{fig:ProbDens4} 
show the energy distribution function $P(E^{\rm intr}_{\rm H};E^{*}_{\rm sc})$
for typical scission shapes for fission of $^{236}$U 
into $^{104}$Zr + $^{132}$Te
and $^{88}$Se + $^{148}$Ce, respectively,
in each case for four different values of the total available energy,
$E^*_{\rm sc}= 5, 10, 20, 30$ MeV.

The distribution functions in Eq.\ \ref{eq:partition}, based on microscopic 
level densities, are compared to the corresponding distribution functions
based on simple Fermi-gas level densities in Figs.\ \ref{fig:ProbDens3}
and \ref{fig:ProbDens4}.
Both types of level density yield rather broad distributions
due to the smallness of the nuclear system.
However, while the Fermi-gas form gives smooth distributions 
that peak where the energy share equals the mass share,
$E^{\rm intr}_{\rm H}/E^*_{\rm sc}=A_{\rm H}/A_0$,
the microscopic level densities lead to distributions 
that exhibit significant irregularities at low total excitation energies.
Furthermore, importantly, at low values of $E^*_{\rm sc}$
the most probable fragment excitation may differ significantly from
the macroscopic expectation given by the Fermi-gas form.
Generally, as the total excitation energy is increased,
these nuclear-structure effects diminish 
and the microscopic energy distribution grows ever smoother
as it gradually approaches the Fermi-gas result.

In the example shown in Fig.\ \ref{fig:ProbDens3},
the heavy fragment, $^{132}$Te, 
is close to being doubly magic and has a large negative shell energy.
It therefore requires a relatively high excitation energy 
to approach the Fermi-gas result.
Furthermore, the low level density of this fragment
causes the light fragment, $^{104}$Zr, 
to be favored in the  energy sharing at most excitation energies. 
For example, when the available total excitation energy is
 $E^*_{\rm sc}$=10 MeV  (Fig.\ \ref{fig:ProbDens3}b), 
on the average about 6 MeV goes to the light fragment $^{104}$Zr 
and only about 4 MeV goes to the heavy fragment $^{132}$Te, 
while the Fermi-gas level densities lead to the reverse energy partitioning.

The other example (Fig.\ \ref{fig:ProbDens4}) 
is a somewhat more asymmetric mass division, $^{88}$Se~+~$^{148}$Ce,
and some favoring of the heavy fragment is apparent, 
in particular at low values of $E^*_{\rm sc}$.  
A quite spectacular situation emerges at the lowest energy shown,
$E^*_{\rm sc}=5$~MeV, where it is predicted that the heavy fragment 
acquires all the energy with a non-negligible probability ($\approx$14\%).
This is partly due to large pairing gaps (thus low level density)
for the light fragment $^{88}$Se, causing the heavy fragment $^{148}$Ce 
to be favored in the energy partitioning.

The excitation energy partition between the heavy and the light fragments 
may thus change significantly with the amount of total excitation energy
available and, furthermore, the partition scenario may change dramatically
from one pair of fission fragments to another. 

\begin{figure}[t]
\centering
\includegraphics[width=1.0\linewidth]{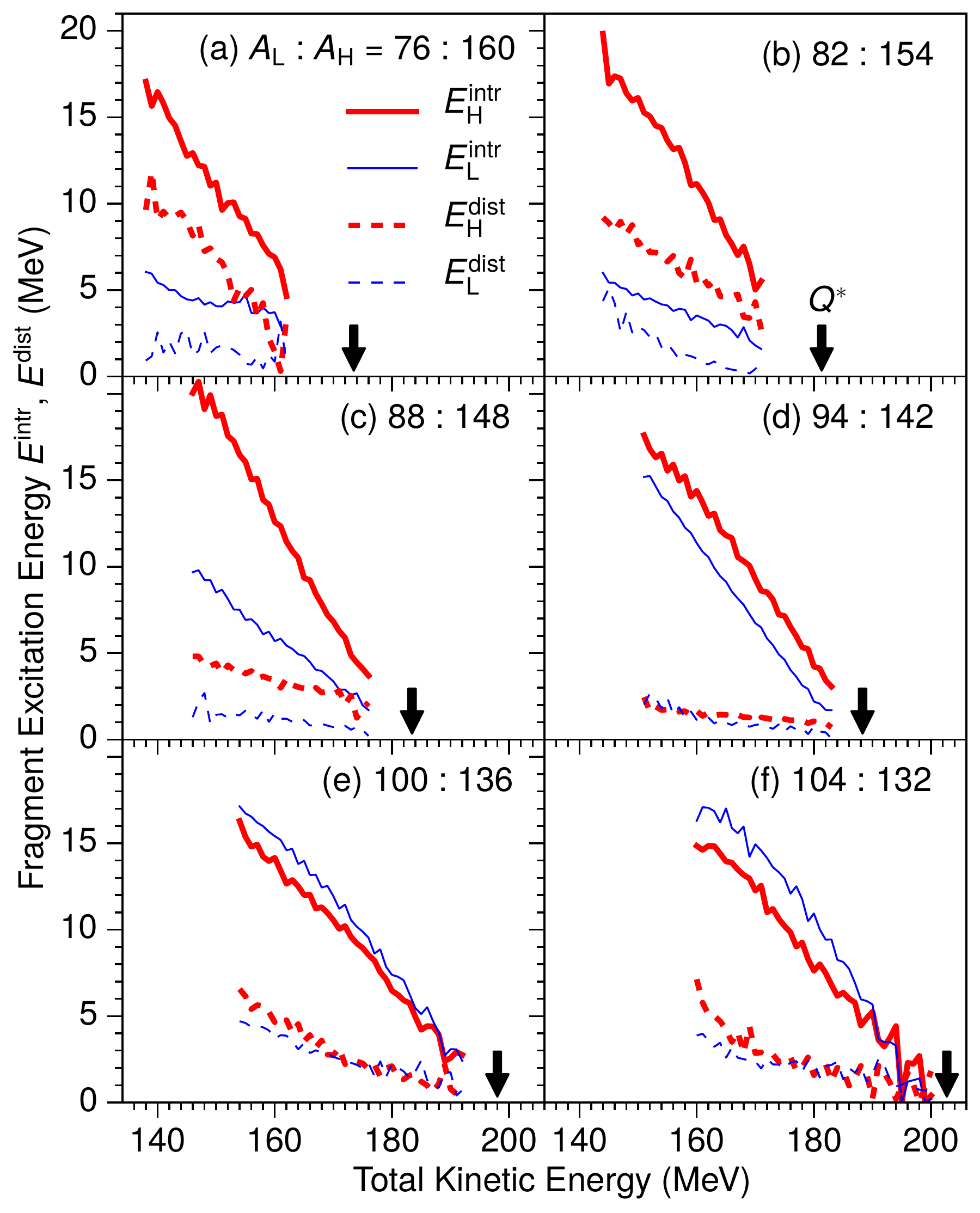}
\caption{
The average of intrinsic energy and distortion energy as functions of the total kinetic
energy for six fragment combinations: (a) $A_{\text{L}}$:$A_{\text{H}}$=76:160, (b) 
82:154, (c) 88:148, (d) 94:142, (e) 100:136, and (f) 104:132, valid for $E_{\rm n}=0$.
The intrinsic energy for the heavy and light fragments is shown
by solid red and blue lines, respectively. Distortion energies are shown by dashed
red and blue lines, respectively. The arrows point to the $Q^*$ values.
}
\label{fig:FragmentEnergies} 
\end{figure}

With this background,
we now consider the average values of the intrinsic excitation energies
as well as the distortion energies of the proto-fragments
resulting from the ensemble of scission configurations
obtained with the Metropolis shape evolution.
Figure \ref{fig:FragmentEnergies} shows the average intrinsic energy 
of the heavy and light fragment versus the specified value of \TKE,
for six selected fragment-pair combinations. 
These examples are chosen to cover the full fragment-mass region considered: 
$A_{\rm L}$:$A_{\rm H}$~= 76:160~(a),
82:154 (b), 88:148 (c), 94:142 (d), 100:136 (e), and 
104:132 (f).

Large nuclear-structure effects are apparent. 
In particular, the large share of intrinsic excitation given to 
the heavy fragment in panel (c) (88:148) 
can be understood from Fig.\ \ref{fig:ProbDens4}, 
and the fact that the light fragment receives the largest energy share
in panel (f) (104:132) can be understood from Fig.\ \ref{fig:ProbDens3},
as discussed above.

\subsection{Distortion energy}
\label{sec:energies_3}

As discussed above, 
the distribution of intrinsic energy in a proto-fragment,
$E^{\rm intr}_i$, varies with the specified value of \TKE.
The fragment distortion energies, $E^{\rm dist}_i$,
are also sensitive to the specified \TKE\
because the sum of the total collective kinetic energy and
the total distortion energy must equal the total available energy
minus the total intrinsic energy,
\begin{equation}
\TKE+E^{\rm dist}_{\rm L}+E^{\rm dist}_{\rm H} = Q^*_{\rm LH}-E^*_{\rm sc}\ .
\end{equation}

\begin{figure}[t]
\centering
\includegraphics[width=1.0\linewidth]{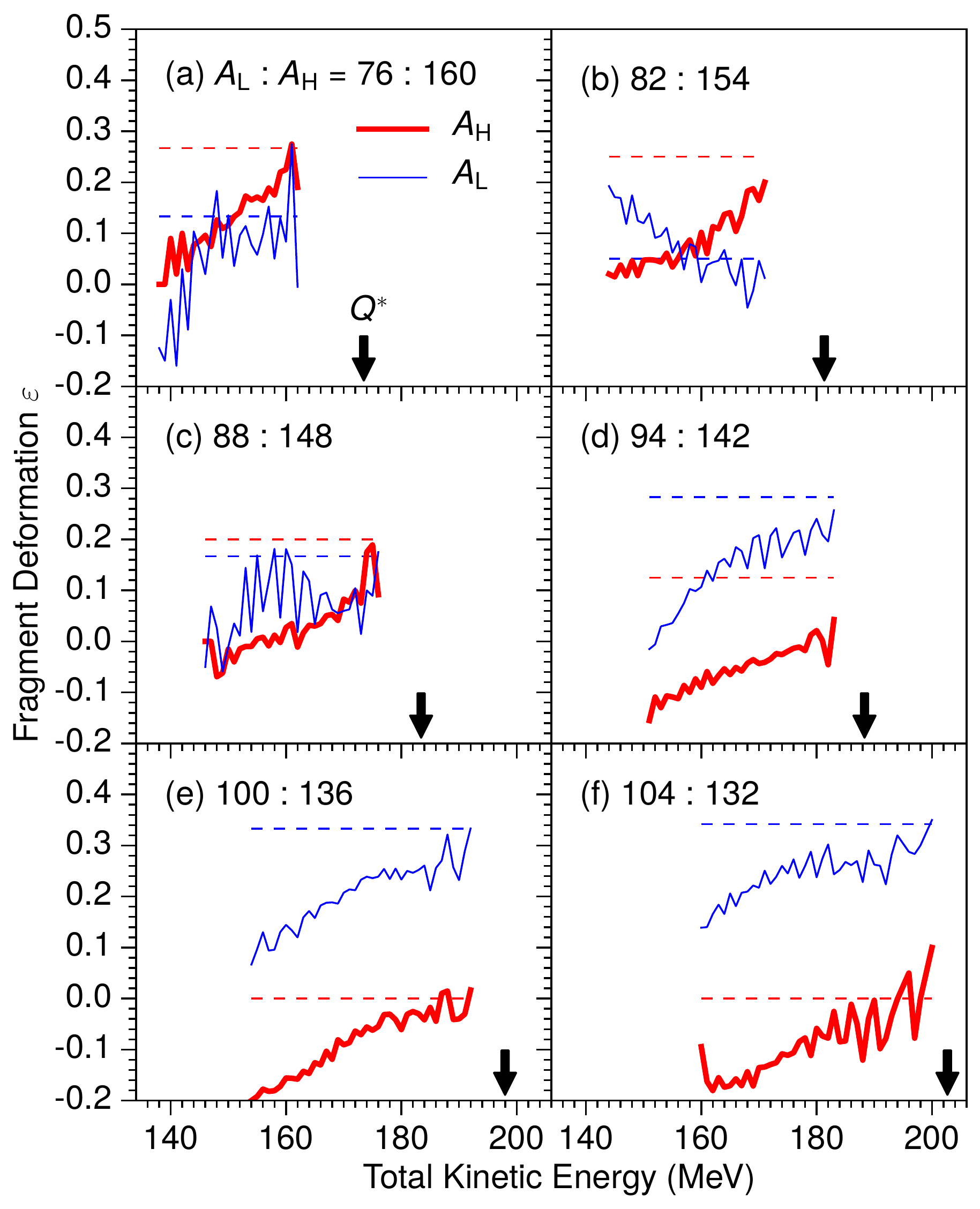}
\caption{
Average fragment deformations at scission versus the total kinetic energy
for the same fragment combinations shown in Fig.\ \ref{fig:FragmentEnergies}.
$E_{\rm n}=0$.
Thick red curves show the heavy fragment deformation and 
thin blue curves the light fragment deformation.
Horizontal dashed red and blue lines show
equilibrium deformations for the heavy and light fragments, respectively. 
The arrows point to the $Q^*$ values. The jaggedness
of the curves is caused by the finite sampling on
a grid of 3QS shapes.
}
\label{fig:FragmentDef} 
\end{figure}

Consequently, at the highest values of \TKE\
the energy balance does not leave much room for fragment distortion.
Conversely, the lowest \TKE\ values are associated with large elongations
of the scission configurations and significant distortions
of the proto-fragments.
The relationship between \TKE\ and the fragment distortion energies
is illustrated in Fig.\ \ref{fig:FragmentEnergies} for six different
mass partitions, and the \TKE\ dependence of the proto-fragment shapes 
is illustrated in Fig.\ \ref{fig:FragmentDef} for the same cases.

The distortion energy resulting from a certain deformation change 
depends strongly on the structure of the specific fragment considered.
For example, for the fragment $^{160}$Sm
a deformation change from the ground-state value $\varepsilon=0.25$ 
to $\varepsilon=0.0$ (Fig.\ \ref{fig:FragmentDef}a) yields a
distortion energy of $E^{\rm dist}=11$~MeV (Fig.\ \ref{fig:FragmentEnergies}a),
while the same deformation change for $^{94}$Kr
(Fig.\ \ref{fig:FragmentDef}d) yields a much smaller distortion energy, 
$E^{\rm dist}=3$~MeV (Fig.\ \ref{fig:FragmentEnergies}). 
This is because the deformation energy curve $U(\varepsilon)$ for $^{160}$Sm
has a robust ({i.e.}\ stiff) prolate minimum,
while that for $^{94}$Kr is rather soft around its prolate minimum
(with respect to both $\varepsilon$ and $\gamma$).

For all mass partitions,
the distortion energy of the heavy fragment is consistently larger 
than that of the light fragment, $E^{\rm dist}_{\rm H}>E^{\rm dist}_{\rm L}$.
This difference grows with increasing mass asymmetry and
the largest difference is found for $A_{\rm L}:A_{\rm H}=76:160$,
while there is almost no difference for 
$A_{\rm L}:A_{\rm H}=94:142, 100:136, 104:132$.

We also note that the distortion energy is systematically smaller 
than the intrinsic energy for given fragment mass partition and \TKE,
$E^{\rm dist}_i<E^{\rm intr}_i$, but the relative contribution from the
distortion energy to the total excitation energy of a fragment 
varies substantially from one fragment to another. 
For example, for $A_{\rm L}:A_{\rm H}=94:142$ both fragments 
receive only 10--15\% from the distortion energy, 
while for $82:154$ the light fragment receives almost half
of its final excitation energy from the distortion energy
at the lowest \TKE\ values,
(but the contribution drops to only about 10\% at the highest \TKE\ values). 

In general, 
our results reveal a quite complex, structure-dependent variation 
of both the intrinsic energy and the distortion energy 
with the fragment identity as well as with \TKE.

\section{Neutron multiplicities}
\label{sec:correlation}
After the primary fission fragments have been fully accelerated
by their mutual Coulomb repulsion and their shapes have relaxed
to their equilibration form,
they typically deexcite by (possibly sequential) neutron evaporation
followed by photon radiation.
The mean number of neutrons emitted from a particular fragment species,
$\nubar_i$, presents a convenient (and observable) measure
of the degree of its initial excitation,
$E^*_i=E^{\rm intr}_i+E^{\rm dist}_i$.

Therefore, in the present study, 
we calculate neutron evaporation from the excited fragments.
We employ the method described in Ref.\ \cite{Randrup2009},
but use the effective microscopic level densities of the daughter nuclei
(rather than the simplified Fermi-gas form),
as in our earlier study \cite{Albertsson2020}.
Thus we assume that the kinetic energy spectrum of 
an evaporated neutron is given by
\begin{equation}
dN_{\rm n}(\epsilon_{\rm n})/d\epsilon_{\rm n}\sim
	\rho'(E'_{\rm max}-\epsilon_{\rm n})\,\epsilon_{\rm n}\ ,
\end{equation}
where $E'_{\rm max}=E^*-S'_{\rm n}$ is the maximum excitation in the
daughter nucleus (corresponding to the evaporation of a neutron
having vanishing kinetic energy $\epsilon_{\rm n}$)
and $\rho'(E')$ is its level density.

For thermal fission, $E_{\rm n}\approx0$,
the average neutron multiplicity from each fragment species
is calculated for specified values of \TKE\ (Sect.\ \ref{sec:TKE})
and for specified values of \TXE\ (Sect.\ \ref{sec:TXE}).
Corresponding results for $E_{\rm n}=5.55$~MeV
are presented subsequently (Sect.\ \ref{sec:Higher}).

\subsection{TKE-gated neutron multiplicities}
\label{sec:TKE}
\begin{figure}[t]
\centering
\includegraphics[width=1.0\linewidth]{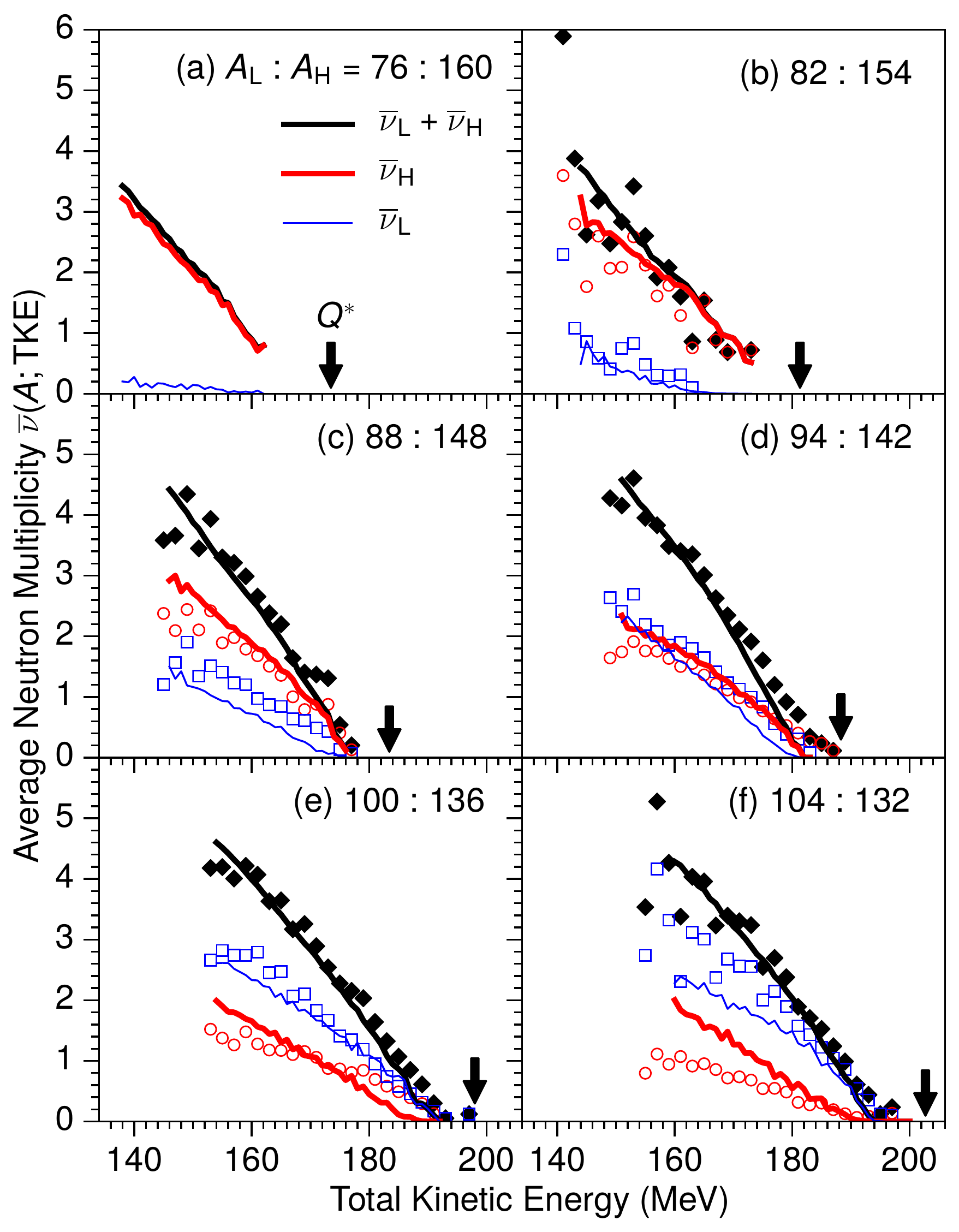}
\caption{
Calculated and measured average multiplicity of neutrons emitted from reaction $^{235}$U(n$_{\text{th}}$,f)
versus \TKE\ for the six fragment combinations of Fig.\ \ref{fig:FragmentEnergies}. 
Measured \cite{Gook2018} light-fragment neutron multiplicity  is shown by red circles, 
heavy-fragment neutron multiplicity by blue squares, and the total is shown by black squares. The calculated 
light-fragment neutron multiplicity is shown by thin blue  lines, the heavy-fragment multiplicity
is shown thick red lines, and the total multiplicity is shown by black  lines.
The arrows point to the $Q^*$ values.
}
\label{fig:NeutronMult} 
\end{figure}

Figure \ref{fig:NeutronMult} shows the calculated average neutron multiplicity
for specified \TKE\ from the light and the heavy fragments, 
$\nubar(A_{\rm L};\TKE)$ and $\nubar(A_{\rm H};\TKE)$, as well as their sum,
for the same six divisions as were considered in
Figs.\ \ref{fig:FragmentEnergies}-\ref{fig:FragmentDef}.
Also shown  (where available) are the experimental results reported by
G\"{o}\"{o}k {et al.}\ \cite{Gook2018}.
In general, the agreement between calculated and measured results 
is very good. 

For very asymmetric divisions the heavy fragment receives 
most of the excitation energy (see Fig.\ \ref{fig:FragmentEnergies}) and,
as a result, it contributes almost all of the neutrons.
This feature is most pronounced for the most asymmetric case displayed,
$A_{\rm L}:A_{\rm H}=76:160$,
but it is also clearly present for $82:154$ and, 
to a smaller degree, for $88:148$.
For $94:142$ the two mean multiplicities are very similar
even though the heavy fragment is $\approx$50\% larger than the light one.
Finally, closer to symmetry ($100:136$ and $104:132$),
where the microscopic effects tend to favor the light fragment 
in the energy division, the neutron multiplicity is dominated
by the light fragment.
This gradual change in the mean neutron multiplicity 
as a function of the mass asymmetry
is also present in the \TKE-averaged results,
as was previously discussed \cite{Albertsson2020}.

The calculated dependence of the total excitation energy
of a given fragment, $E^*_i=E^{\rm intr}_i+E^{\rm dist}_i$,
on the specified value of \TKE\ (Fig.\ \ref{fig:FragmentEnergies}), 
provides an understanding of how the mean neutron multiplicity 
varies with \TKE\ for the various fragment masses. 
For the most asymmetric case considered, $A_{\rm L}:A_{\rm H}=76:160$,
70-80\% of the total excitation energy is carried by the heavy fragment.
Because the light fragment is then typically insufficiently excited
to permit the evaporation of a neutron, practically all of the neutrons
originate from the heavy fragment, for all values of \TKE\
(see Fig.\ \ref{fig:NeutronMult}a).

For several fragment combinations the $\nubar$ dependence
of \TKE\  is almost linear. 
An interesting exception is the threshold effect found for 
the light fragment in the $82:154$ division (Fig.\ \ref{fig:NeutronMult}b),
an effect also seen in the measurements \cite{Gook2018}.
For large values of \TKE\ all neutrons are emitted from the heavy fragment,
while the neutron emission from the light fragment sets in smoothly
at lower \TKE, resulting in a gradual increase of $\nubar_{\rm L}$
from zero to about one as \TKE\ decreases from about 160 to 145 MeV
(Fig.\ \ref{fig:NeutronMult}b). 

For the $88:148$ mass division 
the light fragment is calculated to emit on the average 
somewhat more than one neutron at the smallest \TKE\ values, 
while the heavy fragment dominates,
emitting up to about three neutrons at small \TKE.

For $A_{\rm L}:A_{\rm H}=94:142$ one third of the total excitation energy 
is concentrated in the light fragment at high \TKE, 
and its share increases smoothly with decreasing \TKE\
towards an equal share for both fragments at the smallest \TKE. 
This is seen in Fig.\ \ref{fig:NeutronMult}d:
at high \TKE\ the neutron evaporation is coming only from the heavy fragment,
but as \TKE\ is decreased the ratio changes smoothly leading towards
equal contributions from the light and heavy fragments at small \TKE.
To some degree, this is also seen in the measured neutron multiplicities.

For the two least asymmetric divisions, $100:136$ and $104:132$,
most neutrons are emitted from the light fragment. 
For these cases we have $\nubar_{\rm L}>\nubar_{\rm H}$ for all \TKE\ values,
with the difference decreasing with decreasing \TKE\ for $104:132$.

 For this latter case, the calculations differ somewhat from the data,
with the calculated $\nuH$ exceeding the data for small \TKE\
and the calculated $\nuL$ being smaller than the data.
It is interesting to note the non-linear variation of 
$\nuL$ and $\nuH$ with \TKE,
with $\nuH$ approaching $\nuL$ at small \TKE.
The partition of the intrinsic energy at scission,
which dominates \TXE\ for these fragments, 
provides the major part of the final excitation energy of the light fragment
(see Figs.\ \ref{fig:ProbDens3} and \ref{fig:FragmentEnergies}e). 
But the data suggests that the light fragment's share of the excitation energy
is even larger.
This may be because the dominant fission fragment having $A=132$
is the doubly magic nucleus $^{132}$Sn,
while our requirement of a fixed $N/Z$ ratio for all fragments 
yields $^{132}$Te for $A=132$.
If the doubly magic nucleus $^{132}$Sn is the principal heavy $A=132$ fragment,
then its extraordinarily low level density 
would indeed cause the light fragment to receive
an even larger share of the excitation energy.

\subsection{TXE-gated neutron multiplicities}
\label{sec:TXE}

When the total excitation energy \TXE\ is fixed,
then all different fragment pairs have the same amount of excitation energy 
to share (namely \TXE), and it is instructive to study
the neutron multiplicity $\nubar(A)$ from different fission fragments.
In particular, this would make it possible to investigate 
the variation of $\nubar(A;\TXE)$ as \TXE\ is changed
and study how the sawtooth feature evolves with excitation energy.

It is elementary to obtain $\nubar(A;\TXE)$ from $\nubar(A;\TKE)$
because $\TXE=Q^*_{\rm LH}-\TKE$ for a given light-heavy mass split,
so $\nubar(A;\TXE)=\nubar(A;\TKE=Q_{\rm LH}^*-\TXE)$
where the fragment mass number $A$ determines the L-H division.

\begin{figure}[t]
\centering
\includegraphics[width=1.0\linewidth]{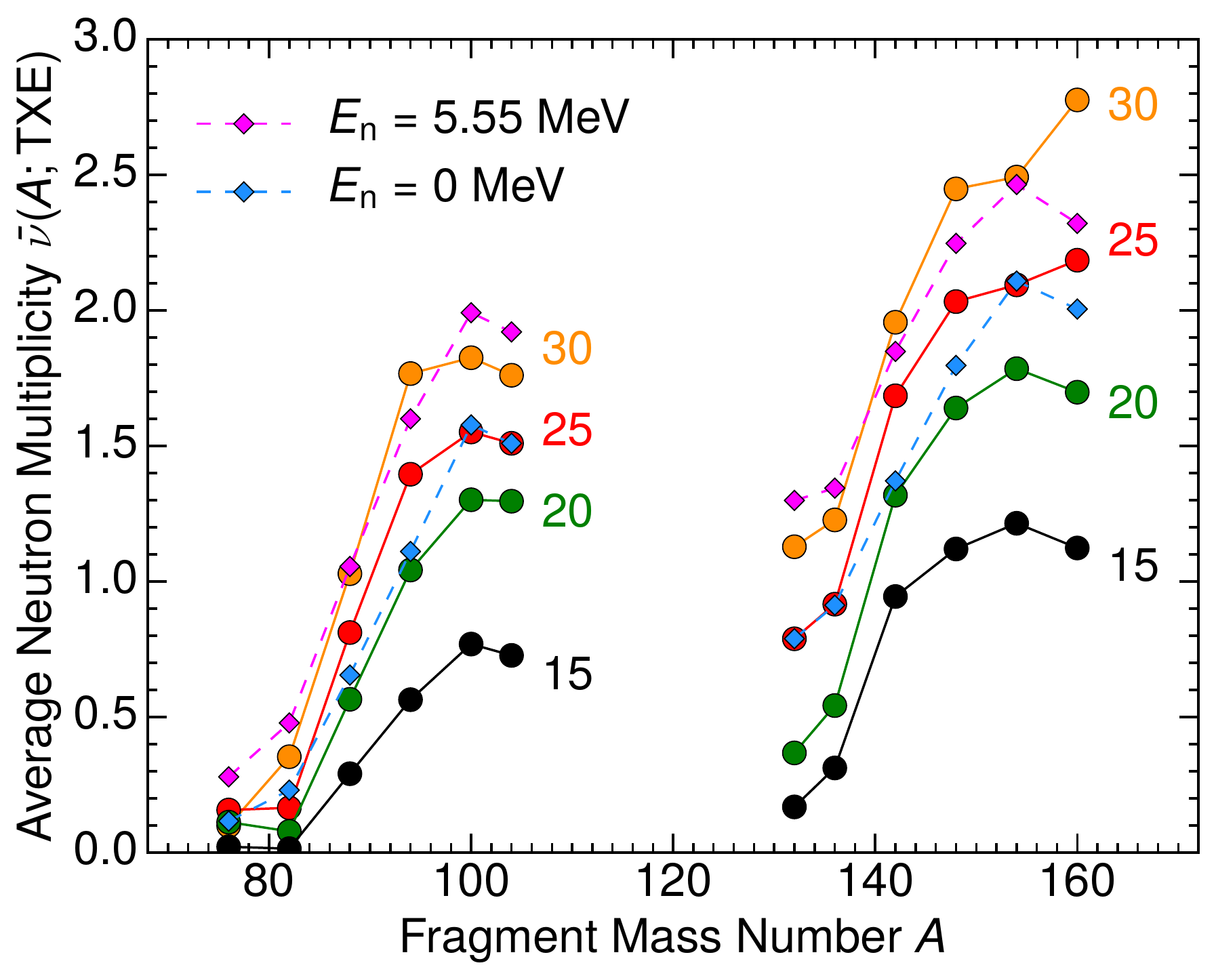}
\caption{
Calculated average neutron multiplicity, $\bar{\nu}(A,\TXE)$, for $^{235}$U$(n_{\text{th}},f)$ for
fixed values of \TXE, 
is shown as function of the fission fragment mass number, $A$. Results are shown for four different \TXE\ values,
\TXE=15, 20, 25 and 30 MeV (filled circles connected by solid lines). The neutron multiplicity averaged 
over \TXE-values (eq.\ \ref{eq:nuAV})
 is shown for $E_{\text{n}}$=0 (blue diamonds connected by dashed lines) and
for $E_{\text{n}}$=5.55 MeV (purple diamonds connected by dashed lines).
 }
\label{fig:AvNeutronMult} 
\end{figure}

Figure \ref{fig:AvNeutronMult} shows $\nubar(A;\TXE)$ for
four different values of the total excitation energy, \TXE=15, 20, 25, 30 MeV.
As discussed in detail above,
the specified total excitation energy \TXE\ is being divided 
quite unevenly between the two fragments
due to the complexity of the separate contributions 
from intrinsic energy and distortion energy.
For all four \TXE\ values,
the average neutron multiplicity from the light fragment, $\nubar_{\rm L}$,
increases with the fragment mass number, $A_{\rm L}$, 
while the multiplicity from the heavy partner fragment decreases 
(except for the cases $A_{\rm L}=104$ and $A_{\rm H}=160$).
At large asymmetries the light fragment evaporates fewer neutrons 
than the heavy fragment,
while the opposite is true closer to symmetry.

With increasing \TXE\ the excitation energy of each fragment increases, 
resulting in larger neutron multiplicities from both fragments. 
However, the increase of the excitation energy of a fragment
is not linear in \TXE, as was discussed 
in Sects.\ \ref{sec:energies_2} and \ref{sec:energies_3}. 
For example, for $A_{\rm L}:A_{\rm H}=104:132$
an increase of \TXE\ by 5 MeV from 15 to 20 MeV 
results in an energy increase of about 3.5 MeV in the light fragment 
and 1.5 MeV in the heavy fragment,
leading to multiplicity increases of about $\Delta \bar{\nu}_{\rm L}$=0.6 
and $\Delta\nubar_{\rm H}$=0.2 (see Fig.\ \ref{fig:AvNeutronMult}).
By contrast, the same increase in \TXE\ from 25 to 30 MeV 
causes an energy increase of about 2.5 MeV in both fragments
leading to $\Delta\nubar_{\rm L}$=$\Delta\nubar_{\rm H}$=0.3. 
This evolution away from light-fragment dominance
with increasing \TXE\ is caused primarily by the specific
energy-dependence of the level densities in the distorted proto-fragments
which reduces the favoring of the light fragment 
in the sharing of the intrinsic energy at scission,
as is seen in Figs.\ \ref{fig:ProbDens3}b and \ref{fig:ProbDens3}c.

The neutron multiplicity from more asymmetric divisions
show a less dramatic evolution with \TXE.
For example, for $A_{\rm L}:A_{\rm H}=82:154$ the multiplicity increases 
are $\Delta\nubar_{\rm L}$=0.1 and $\Delta \nubar_{\rm H}$=0.6
when \TXE\ is increased from 5 to 10 MeV,
and  $\Delta\nubar_{\rm L}$=0.2 and $\Delta\nubar_{\rm H}$=0.4
when \TXE\ is increased from  25 to 30 MeV.

When \TXE\ is increased from 15 to 30 MeV for $A_{\rm L}:A_{\rm H}=76:160$ almost all increase in
excitation energy goes to the heavy fragment. This results 
in the large increase in heavy-fragment neutron multiplicity, $\Delta \nubar_{\rm H}$=1.7,
as compared to only $\Delta \nubar_{\rm L}$=0.3 for the light fragment.

Figure \ref{fig:AvNeutronMult} also shows the unconstrained neutron
multiplicity, $\nubar(A)$, for both thermal fission, $E_{\rm n}\approx 0$,
and for the higher energy considered in Sect.\ \ref{sec:Higher},
$E_{\rm n}=5.55$~MeV.
The unconstrained multiplicity can be regarded as a weighted average
of the \TXE-constrained multiplicity,
\begin{equation}
\nubar(A)={\int{N(A;\TXE)\,\nubar(A;\TXE)\,d\TXE}\over\int{N(A;\TXE)\,d\TXE}}\ ,
\label{eq:nuAV}
\end{equation}
where $N(A;\TXE)$ denotes the number of events leading to the specified
value of \TXE.
For thermal fission the average \TXE\ value is found to be 
around 25 MeV for most the $A$ values shown, 
except for $A_{\rm L}:A_{\rm H}=88:148$ and $94:142$
where $\overline{\TXE}=21-22$~MeV.
We note that $\nubar(A)$ agrees very well with $\nubar(A;\overline{\TXE})$
for both thermal fission and for $E_{\rm n}$=5.55 MeV
for which we have $\overline{\TXE}\approx30$ MeV.

\subsection{Higher neutron energies}
\label{sec:Higher}
We also discuss our results for a higher neutron energy, 
$E_{\rm n}=5.55$~MeV. This neutron energy was
previously considered in experiments \cite{MullerPRC29}, and is close
to the maximal energy for first-chance fission.

\begin{figure}[t]
\centering
\includegraphics[width=1.0\linewidth]{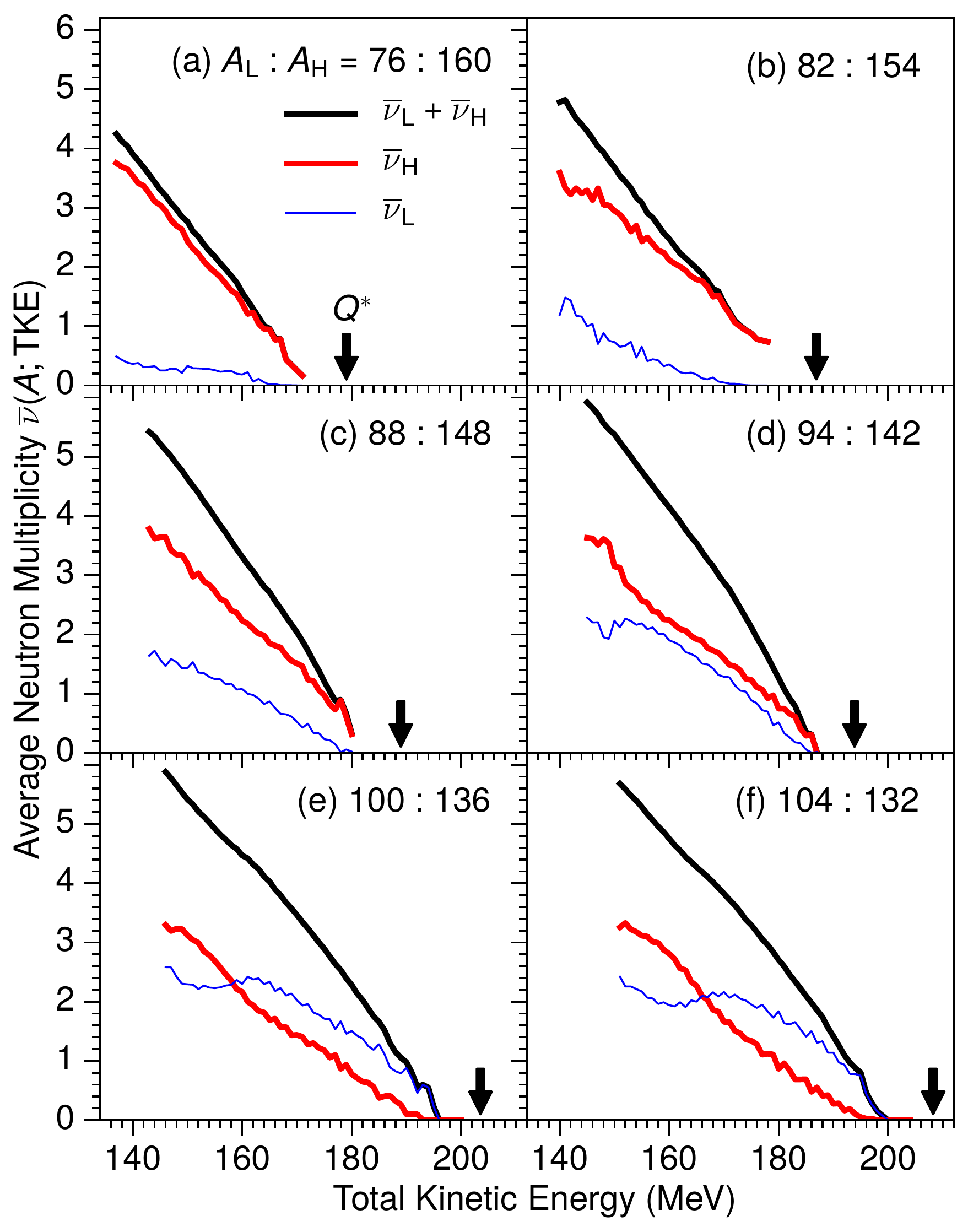}
\caption{
Same as Fig.\ \ref{fig:NeutronMult}  but for $E_{\text{n}}$=5.55 MeV.
 }
\label{fig:NeutronMult2} 
\end{figure}

Figure \ref{fig:NeutronMult2} shows the \TKE-gated mean neutron multiplicity,
$\nubar(A;\TKE)$ for the same six mass divisions as studied above.
When the energy of the incoming neutron increases,
the $Q$ value increases correspondingly because the initial excitation
energy of the fissioning nucleus is $E_0^*=S_{\rm n}+E_{\rm n}$.
Thus \TKE+\TXE\ is increased by $E_{\rm n}$
and we find that the average \TKE\ changes very little
 and most of the additional energy goes to \TXE.
As a consequence, the shape evolution is able to explore a wider
domain of the potential-energy landscape and the system gains access
to valleys that lead to more elongated scission shapes.
This feature will be reflected in a bimodal character of the
\TKE\ distribution (see Sect.\ \ref{bimodal}).

For the three most asymmetric divisions, Figs.\ \ref{fig:NeutronMult2}a-c,
the behavior of $\nubar(A;\TKE)$ is similar to the thermal result
for both $\nubar_{\rm L}$ and $\nubar_{\rm H}$
(see Figs.\ \ref{fig:NeutronMult}a-c),
except for an overall increase 
due to the increased excitation of the primary fission fragment.
On average, the neutron multiplicity increases 
by 0.3-0.4 for both the light and the heavy fragments.

This smooth evolution with $E_{\rm n}$ may be contrasted with the behavior
for the less asymmetric divisions (Figs.\ \ref{fig:NeutronMult2}d-f)
where qualitative changes are apparent.
For the two least asymmetric cases, it is especially noticeable 
that $\nubar(A_{\rm L};\TKE)$ and $\nubar(A_{\rm H};\TKE)$ cross 
so the heavy fragment becomes dominant at low \TKE,
with the crossings occuring at \TKE\ values of 160 and 164 MeV
for $100:136$ and $104:132$, respectively.
Such an evolution would be expected from the fact that the
increased intrinsic excitation of the proto-fragments tends to wash out 
the structure effects that favored the light fragment. 
In the next Section we discuss how the appearence of a new superlong 
fission mode plays an important role for this behavior.

\subsubsection{Bimodal fission}
\label{bimodal}

\begin{figure}[t]
\centering
\includegraphics[width=1.0\linewidth]{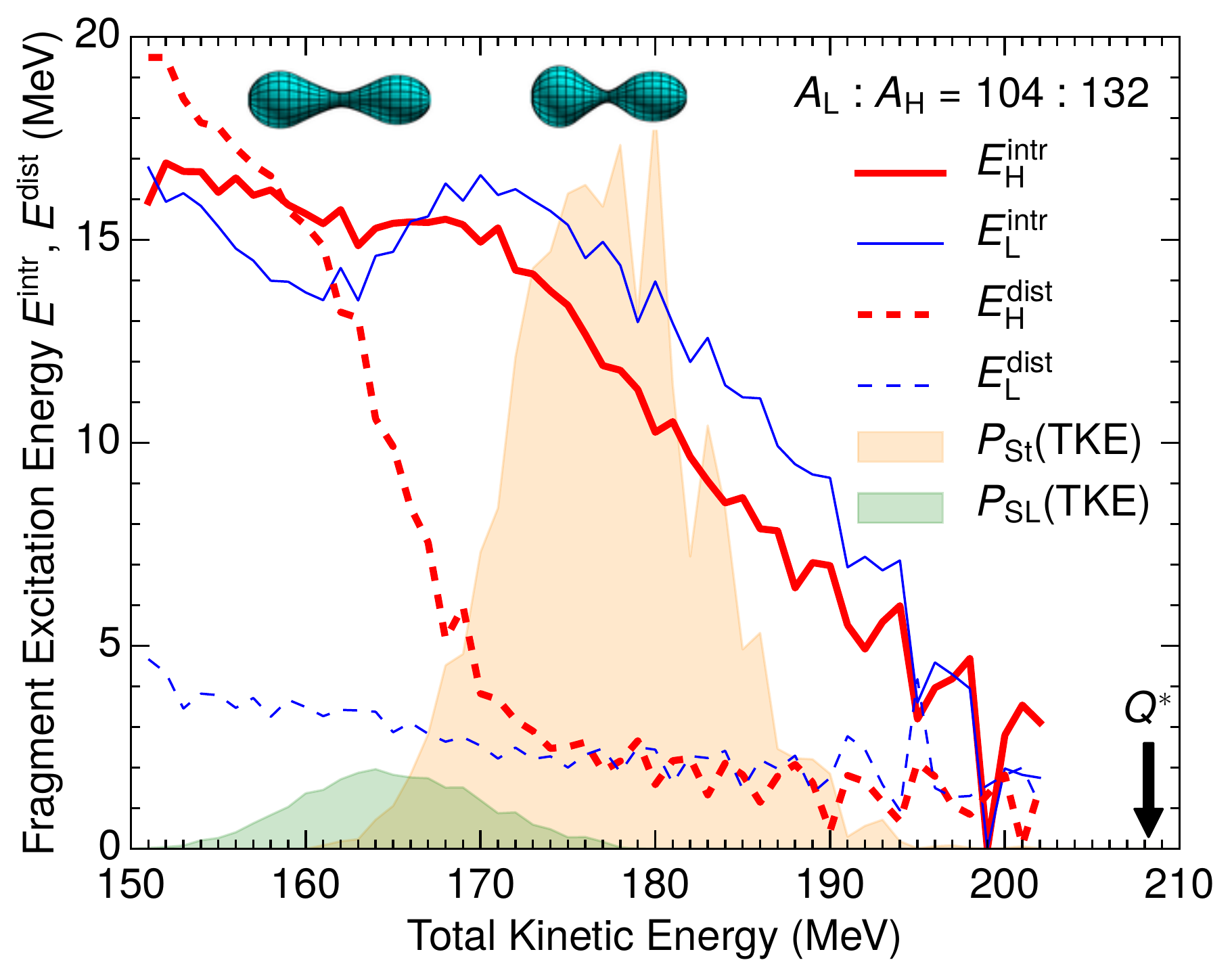}
\caption{
Similar to Fig.\ \ref{fig:FragmentEnergies}f, with fragment mass ratio
$A_{\text{L}}$:$A_{\text{H}}$=104:132, for $E_{\text{n}}$=5.55 MeV.
Probability densities are shown for the two modes: superlong (SL), with average
shape specified by $q_2$=16.5, $\varepsilon_{\text{L}}$=0.15, $\varepsilon_{\text{H}}$=0.32
(green area), and standard (St) with average shape
$q_2$=11.5, $\varepsilon_{\text{L}}$=0.25, $\varepsilon_{\text{H}}$=-0.08 (orange area). 
Typical shapes of the SL and St modes are shown.}
\label{fig:Energies2} 
\end{figure}

Figure \ref{fig:Energies2} shows the separate contributions 
to the final fragment excitation energy 
from their intrinsic and distortion energies at scission,
displayed versus the resulting \TKE\ for $A_{\rm L}:A_{\rm H}=104:132$.
As mentioned above, the increase in the kinetic energy of the incoming neutron
primarily causes the intrinsic energy to increase
and that in turn gives the system access to 
a wider variety of shapes during its evolution.
This results in the appearance of a new fission mode
characterized by more elongated scission shapes and,
consequently, lower \TKE\ values, see Fig.\ \ref{fig:q2vsTKE}b. 
On average, the scission shapes of these events have $q_2\approx16.5$
significantly larger than those reached  with thermal neutrons
(Fig.\ \ref{fig:q2vsTKE}a).

The existence of such a {\it superlong} (SL) fission mode
has long been known \cite{Brosa1990}. The mode
favours symmetric fission, and it is believed that
the observed increase of emitted neutrons near symmetry with incident
neutron energy is caused by an increase in the yield of the 
superlong mode, see e.g.\ Refs.\  \cite{Hambsch2003,Schmidt2010}.

For the large elongations characterizing the superlong mode 
it is preferable for the heavy fragment 
to develop a large quadrupole deformation
and in average we have $\varepsilon_{\rm H}$=0.32. 
For small \TKE\ values the average quadrupole moment of the heavy fragment 
is even larger, for example $\varepsilon_{\rm H}\approx0.5$ at \TKE=150 MeV.
Also the light fragment is deformed, 
but with a smaller deformation, $\varepsilon_{\rm L}\approx0.15$.

The scission shapes of the SL mode are compared in Fig.\ \ref{fig:q2vsTKE}b 
with those obtained at higher \TKE\ values
where the elongation is much smaller, $q_2 \approx 10-12$. 
In the normal mode, the shape of the heavy fragment is slightly oblate, 
$\varepsilon_{\rm H} \approx -0.08$ 
while the light fragment has $\varepsilon_{\rm L}\approx0.15$.
We refer here to this fission mode as {\it standard} (St).

Figure \ref{fig:Energies2} also shows calculated distributions 
of the two fission modes versus \TKE.
Guided by the result shown in Fig. \ref{fig:q2vsTKE}b,
the SL mode is defined by the condition
$q_{2}>14$ and correspondingly the St mode is defined by $q_{2}<14$.
The St mode exhibits a broad distribution with its maximum at 
$\TKE\approx180$ MeV and completely dominates the fission process at high \TKE.
But with decreasing \TKE\ values the SL mode gradually appears, 
at first partly overlapping with the St mode 
but then taking completely over in the lowest \TKE\ range.

The maximum of the SL distribution occurs at $\TKE\approx164$ MeV. 
This average \TKE\ value for the SL mode, converted from fragment to product kinetic energy
becomes $161$ MeV, and is in reasonable agreement 
with the measured product kinetic energy for mass split $A_{\rm L}:A_{\rm H}=104:132$, namely 
156 MeV \cite{Yanez2018}. 
Also the calculated average \TKE\ value for the
St mode, 178 MeV, converted to the product kinetic energy 176 MeV,
is in good agreement with the corresponding measured 
value 174 MeV.
Often \cite{Brosa1990,Yanez2018}, the St mode is described as consisting of two modes,
S1 and S2, which together account for the asymmetric fission, and the value 174 MeV
is a weighted average over these modes from the results given in Ref.\ \cite{Yanez2018}.

The SL mode has maximal influence for symmetric fission but is increasingly involved
in more asymmetric fission splits with increasing neutron energy. For thermal
neutrons the SL mode gives zero contribution to fission events with mass split
 $A_{\rm L}:A_{\rm H}=104:132$, while the SL mode contributes
about 10\% of the events at $E_{\rm n}=5.55$ MeV, see Fig.\ \ref{fig:Energies2}. This agrees fairly well
with data where the SL mode  is found to contribute with 0\% and 5\% at $E_{\rm n} \approx 2$ MeV 
and $E_{\rm n} \approx 5$ MeV, respectively \cite{Yanez2018}.

The large deformations of the heavy fragment in the SL mode 
implies very large distortion energies, $E^{\text{dist}}_{\rm H}$=15-20 MeV 
for \TKE=150-160 MeV, 
because the fragment shape has to relax from $\varepsilon_{\rm H}=0.4-0.5$
to its spherical ground-state shape. 
For comparison, the distortion energy of the heavy fragment 
in the St mode is only 2-3 MeV.

It is interesting to study how the partition of the intrinsic energy 
between the light and heavy fragments depends on the fission mode
for mass division around $A_{\rm L}$:$A_{\rm H}$=104:132.
In the St mode the heavy fragment has a small oblate deformation, 
close to its doubly magic spherical ground state.
The spherical shell gaps at $Z=50$ and $N=82$ 
lead to a very low level density for the heavy fragment,
causing the light fragment to be favored, see Fig.\ \ref{fig:ProbDens3}. 
On the other hand, in the SL mode the heavy fragment
has a substantial quadrupole deformation. 
The proto-fragment is thus far away from equilibrium and has a large
single-particle level density near the Fermi level,
leading to a large total level density.
The distribution function for the intrinsic energy at scission,
Eq.\ (\ref{eq:partition}), then somewhat favors the heavy fragment.

Relative to the St mode,
the heavy fragment receives significantly more excitation energy in the SL mode
due to two different mechanisms:
First, as just discussed, 
it is the favored recipient of intrinsic energy at scission.
Second, as also mentioned above, it is very distorted at scission
and its shape relaxation leads to an additional significant contribution.

The increase of the neutron multiplicity from the heavy fragment with
increasing incident energy is thus partly due to the appearance of the SL mode. 
For thermal fission only the St mode appears and $\nubar_{\rm L}$ 
is larger than $\nubar_{\rm H}$ for all \TKE. 
With increasing neutron energy the SL mode appears 
and causes the heavy fragment to become preferentially more excited.
Consequently, with increasing $E_{\rm n}$,
$\nubar_{\rm H}$ increases faster than $\nubar_{\rm L}$.

Although  the above detailed analysis was carried out for the specific division
$A_{\rm L}$:$A_{\rm H}$=104:132, it is expected to hold also for the
neighbouring mass divisions, as is suggested by
Fig.\ \ref{fig:NeutronMult2}e for $A_{\rm L}$:$A_{\rm H}$=100:136.
For this mass split we find about 5\% SL mode at $E_{\rm n}$=5.55 MeV
that can be compared to data that gives 3\% \cite{Yanez2018}.
With the smaller contribution of the SL mode 
the dominance of $\nubar_{\rm H}$ over $\nubar_{\rm L}$
sets in at a lower
\TKE\ value for the mass split $A_{\rm L}$:$A_{\rm H}$=100:136 than
for $A_{\rm L}$:$A_{\rm H}$=104:132 (Fig.\ \ref{fig:NeutronMult2}e,f).

Indeed, it has been found experimentally \cite{MullerPRC29}
that the additional prompt neutrons emitted when the incident neutron energy 
is increased originate mainly from the heavy fragment.
The onset of the SL mode with increasing neutron
energy provides an additional mechanism for the neutron 
multiplicity from the heavy fragment to increase more than from the light fragment.

\section{Summary and discussion}
\label{sec:summary}

Based on macroscopic-microscopic potential-energy
surfaces in the 3QS shape parametrization, and microscopic
level densities, we have applied the Metropolis random walk method
to treat the induced fission reaction $^{235}$U(n,f)
at $E_{\text{n}}$=0 and 5.55 MeV.
Because the calculational method generates an ensemble of individual
fission events it is possible to extract a large variety of correlations.
We have particularly studied $\nubar(A;\TKE)$, the average neutron multiplicity
as function of the primary fission-fragment mass number $A$
for events leading to a specified total fragment kinetic energy \TKE.
For a given mass division, $A_0\to A_{\rm L}+A_{\rm H}$,
the \TKE\ constraint restricts the total combined fragment excitation energy
to be $\TXE=Q_{\rm LH}^*-\TKE$.
The excitation of each fragment is composed of two terms:
its share of the total available intrinsic excitation energy at scission
and the energy recovered from the relaxation of its distorted shape 
of the proto-fragment.
The division of the intrinsic excitation energy between the proto-fragments
is carried out statistically based on their microscopic level densities.
As a consequence, contributions to each fragment from 
the intrinsic excitation energy and the distortion energy
vary with $A$ and \TKE\ in a non-trivial manner.

The indirect observation of the excitation energy 
in each primary fragment is made through the neutron multiplicity. 
For thermal fission, where \TKE-gated data is available, 
a good agreement is obtained between the calculated and the measured 
$\nubar(A;\TKE)$.
This agreement extends to a number of finer details,
such as the threshold effect in neutron multiplicity 
from the light fragment with decreasing \TKE\
in fission leading to the division $A_{\rm L}:A_{\rm H}=82:154$.
Certain differences between calculated and measured neutron multiplicity 
versus \TKE\ for $A_{\rm L}:A_{\rm H}=104:132$
may be due to the fact that the calculation assumes that
all the fragments have the same $N/Z$ ratio.
The inclusion of the fragment isospin degree of freedom
would probably make $^{132}$Sn the most favored fragment,
rather than the neighboring $^{132}$Te,
and, consequently, increase the difference between 
$\nubar(A_{\rm L};\TKE)$ and $\nubar(A_{\rm H};\TKE)$,
as is experimentally observed.

We also studied $\nubar(A;\TXE)$,
the mass-dependent average neutron multiplicity for events
having the specified value of \TXE,
the combined excitation energy of the two primary fragments.
This allows more detailed studies of the sawtooth behavior of
$\nubar(A)$ with excitation energy.

An increase of the incident neutron energy to $E_{\rm n}=5.55$ MeV 
leads to the appearance of a new superlong fission mode
characterized by low \TKE\ values
and occurring in the near-symmetric region,
in agreement with experimental findings.
In the SL fission events,
which involve very elongated scission configurations,
the heavy proto-fragment is particularly distorted, implying
a large distortion energy, but also a high level density.
For these reasons, most of the additional energy brought in by the
neutron goes to excitation of the heavy primary fragment.
Consequently, $\nubar_{\rm H}$ increases faster than $\nubar_{\rm L}$
with increasing $E_{\rm n}$, as is also observed experimentally.

The appearance of the SL mode at higher neutron energies
for the fission mass divisions $A_{\rm L}:A_{\rm H}=104:132$ and $100:136$
is seen in the calculated correlated neutron multiplicity: With decreasing \TKE,
$\nubar(A_{\rm L};\TKE)$ and $\nubar(A_{\rm H};\TKE)$ cross 
so the heavy fragment grows dominant at low \TKE\ (Fig.\ \ref{fig:Energies2}e,f).
This prediction could be tested experimentally with correlation data
obtained for higher neutron energies.

The region of symmetric fission was excluded in our present study
because it appears that the 
scission configurations encountered for near-symmetric divisions
are associated to too large \TKE\ values. This may be because 
the employed 3QS shape parametrization
is not sufficiently flexible to accommodate the rather elongated
scission configurations.
We are currently investigating this problem
and hope to develop a suitable refinement.

Other models of the fission process, such as those employed in
Refs.\ \cite{Dubray2008,ScampsNature2018,BulgacPRL116,Bulgac2018},
suggest larger distortions of the proto-fragments
than what was found in the present treatment.
Those models may nevertheless be able to also give a reasonable
reproduction of the measured neutron multiplicities
if they compensate for the large distortion energies
by giving correspondingly smaller intrinsic energies.
In order to elucidate the situation,
it would be very valuable to calculate within those models as well
the \TKE-constrained neutron multiplicities, $\nubar(A;\TKE)$.
Detailed comparisons of this observable,
both between the various models and with the experimental data,
might reveal the quantitative importance of the different contributions
to the fragment excitations and thus help to improve our understanding 
of the fission process.

\begin{acknowledgments}
We acknowledge discussions with C. Schmidt, and thank  A. G\"{o}\"{o}k and
W. Loveland for providing information about the 
experimental data. Comments on the manuscript by R. Vogt are acknowledged.
This work was supported by the Swedish Natural Science Research Council (S.{\AA}.) and the Knut and Alice
Wallenberg Foundation (M.A., B.G.C. and S.{\AA}.); J.R. was supported in part by the NNSA DNN R\&D of the U.S. Department of Energy.
\end{acknowledgments}

\end{document}